

\input jnl

\line{\hfil NSF-ITP-92-13}
\line{\hfil hepth@xxx/9202075}


\let\ep=\varepsilon

\let\si=\sigma

\let\om=\omega

\let\<=\langle
\let\>=\rangle

\let\txt=\textstyle

\def\e{ {\rm e} }

\def\Dsl{\,\raise.15ex\hbox{/}\mkern-13.5mu D}

\def\comment#1{ \hbox{Comment suppressed here.} }

\pretolerance=10000  
\hbadness=1500  
%
%
\def\natspace{\nulldelimiterspace 0pt \mathsurround=0pt}

\def\Big#1{{\hbox{$\left#1\vbox to11.0pt{}\right.\natspace$}}}

%

\def\DDD{I\kern -0.35em D\,}
\def\RRR{I\kern -0.35em R\ }
\def\nnn{/\kern -0.60em \nabla\ }

\title            S-Wave Scattering of Charged Fermions
                   by a Magnetic Black Hole

\author           Mark G. Alford
                  ~~
                  {\it Institute for Theoretical Physics}
		  {\it University of California}
                  {\it Santa Barbara, CA 93106}
                  ~~
                  and
                  ~~
                  Andrew Strominger
                  ~~
                  {\it Department of Physics}
		  {\it University of California}
                  {\it Santa Barbara, CA 93106}
                  andy@denali.physics.ucsb.edu

\abstract
We argue that, classically,
$s$-wave electrons incident on a magnetically
charged black hole are swallowed with probability one: the reflection
coefficient vanishes. However, quantum
effects can lead to both electromagnetic and gravitational backscattering.
We show that, for the case of extremal, magnetically
charged, dilatonic black holes and a single flavor of low-energy
charged particles,
this backscattering is described by a perturbatively computable and
unitary $S$-matrix, and that the Hawking radiation in these modes
is suppressed near extremality.
The interesting and much more difficult case of several flavors is
also discussed.
\endpage
To many physicists it has seemed natural that extremal black holes,
stabilized by their charge, should interact with their surroundings
very much as elementary particles. In particular,
they should scatter incident particles in a manner described by a
unitary, quantum-mechanical $S$-matrix. This contrasts with
another view, according to which the black hole absorbs
a particle and then Hawking radiates[\cite{HAW}] back to its
extremal state. The Hawking radiation is in a mixed state, so
there is no $S$-matrix.
In support of the ``elementary particle'' picture, it has
recently been shown[\cite{GHS,hw,dynam}] that in many cases potential
barriers arise around near-extremal dilatonic black holes[\cite{GIMA,GHS}],
which
reflect incident particles before they reach the horizon.
In this paper we examine the threat posed to this picture by
an exception to that result:
the $s$-wave charged fermions in the presence of a
magnetic black hole, which classically sees no barrier.
We show that these modes are reflected from
extremal ($a=1$) dilatonic black holes by an
infinite quantum-mechanical barrier, and hence $s$-wave fermion-hole
scattering is indeed described by a unitary $S$-matrix.

Let us first
consider the classical theory consisting of gravity, electromagnetism and
a massless left-handed ``electron''.  Of course
this theory is anomalous at the quantum level, but this does not prevent
us from studying classical scattering solutions of the Dirac equation.
In particular we are interested in the scattering of charge ${e}$
electrons from a black hole of magnetic charge ${g}$ such that ${eg} =
{1\over 2}$.

In the study of wave equations in a fixed black hole background,
one expands the field $\Phi$ into modes $\Phi(\ell,\omega)$
of energy $\omega$ and
angular momentum $\ell $. The wave equation for a bosonic mode $\Phi
(\ell,\omega)$ may then be written in the general form
$$
\biggl(-{\omega^2} + \Bigl({{\partial}\over{{\partial}{r^*}}}\Bigr)^2 +
{V}(\ell )\biggr)\Phi(\ell,\omega) = 0\eqno(md)
$$
where ${r^*}$ is a convenient radial coordinate. For
fermions there is a similar first order differential equation. The
presence of a non trivial potential ${V}$ -- in general non-vanishing even for
$\ell =0$ -- implies the possibility of backscattering. Some component
of a matter $s$-wave impinging on a black hole will in general
backscatter off the gravitational field, while the rest will continue on
across the event horizon.

However something rather special can occur for an electrically charged
fermion in the presence of a magnetically charged black hole.  The
total angular momentum
$$
{\vec J} = {\vec s} + {\vec\ell}+{\hat r}{eg}\eqno(jslf)
$$
\noindent contains spin $({\vec s})$, orbital $({\vec\ell})$ and field
$({eg}{\hat r})$ contributions. If ${eg} = {1\over 2}$ (or more generally
$n+\half$), then there
is a ${\vec J}=0$ $s$-wave mode for which the spin and field
contributions cancel. (These are of course the Callan-Rubakov modes
which lead to monopole catalysis of proton decay [\cite{rub, cal}].)

An incoming, ${\vec\ell}=0$, left handed electron has
its spin aligned with its (radial) momentum, so that ${\vec J} = 0$;
but in a theory with only left-handed electrons, there
are {\it no} outgoing modes with ${\vec J} = 0$. Therefore angular
momentum conservation forbids an incoming electron from
backscattering off of a black hole. This will continue to be
true classically, even if we include the other chirality and allow
Dirac mass terms. We conclude that an
incoming fermion will continue across the horizon and into the
singularity with unit amplitude, and the analog of the
potential $V$ must therefore vanish. In reference [\cite{hw},\cite{alf}]
this was confirmed by explicit calculation, and in [\cite{hw}] it was
pointed out that the absence of a potential barrier would
seem to contradict the ``elementary particle'' picture of
extremal black holes, and allow catastrophic Hawking radiation
in these modes (at least
in the regime a little above extremality, where
back-reaction can be neglected).

Quantum mechanically the situation differs. We are forced to
include both chiralities in order to avoid anomalies. There are then
outgoing ${\vec J} = 0$ fermion modes. If backreaction is included,
as the $s$-wave fermion
impinges on the black hole large electric fields are produced in its
wake. These large electric fields are unstable due to
Schwinger pair production. This is a
quantum mechanical form of backscattering. It is
essentially an electromagnetic effect, and is distinct from Hawking
emission.

This problem is difficult to analyze in general. We shall consider only
a special case which turns out to be particularly simple: the backscattering
of an $s$-wave electron incident on an extremal, magnetic charge $Q$
dilatonic black hole[\cite{GIMA},\cite{GHS}].
These black holes are extrema of the action:
$$
S = \int d^4x\,\sqrt{-g}
\biggl( {\e^{-2\phi}} \Bigl( {R}+{4}(\nabla\phi)^2
-{1\over 2}{F^2}\Bigr) +i {\bar\psi}\Dsl\psi\biggr)\eqno(fouract)
$$
where $\phi$ is the scalar dilaton field, $\psi$ is a charged fermion and
$D$ the covariant derivative\footnote{*}{The power of
$e^\phi$ appearing in front of the fermionic part of the action is
irrelevant
since it may be eliminated (modulo quantum anomalies which could be
important for gravitational backreaction) by a rescaling
of $\psi$.}.
A key feature of this type of black hole is the following. As pointed out
in [\cite{GHS}] and described in detail in [\cite{dynam}], the geometry
consists of three regions. The first is the asymptotically flat (AF) region
far from the black hole. Nearer the black hole the curvatures begin to
rise and one enters the mouth region. The mouth leads into an
infinitely long throat region. Well into the throat region, the metric
is approximated by the flat metric on two-dimensional Minkowski space
times the round metric on the two-sphere with radius $Q$ (in Planck units).
The spacetime has no horizons or singularities.  The dilaton
field $\phi$ increases linearly with the proper distance into the
throat. The electromagnetic field strength is tangent to and
integrates to $4 \pi Q$ over the two-sphere.

At scales large relative to the radius $Q$ of the two-sphere,
dynamics within the throat region are described by a
two-dimensional effective action [\cite{ban},\cite{dynam}]. Regarding
the two-sphere
as a ``compactification manifold'' this effective action can be
derived with standard Kaluza-Klein technology. The
result is [\cite{ban},\cite{dynam}]:
$$
S = \int d^2\sigma\,\sqrt{-g}
\biggl( {\e^{-2\phi}} \Bigl( {R}+{4}(\nabla\phi)^2 +{1\over{Q^2}}
-{1\over 2}{F^2}\Bigr) + i{\bar\psi}\Dsl\psi\biggr)\eqno(twoact)
$$
where ${g},\phi$
and ${F} = {dA}$ are relics of four dimensional fields taking constant
values on the two-sphere and the charged Dirac fermion $\psi$ results from a
zero mode of the charged Dirac equation on a two-sphere threaded
with magnetic flux. The four-dimensional
extremal black hole corresponds to the two-dimensional linear
dilaton vacuum:
$$\eqalign{
{ds^2}&={-}{d}{\tau^2}+{d}{\sigma^2}\cr
\phi(\si)&={\phi_0}+\sigma/2Q\cr
{F}&=0.\cr}\eqno(ldv)
$$
As described in some detail in [\cite{ban},\cite{dynam}],
we can describe the the process of scattering
off an extremal black hole by choosing some value of $\si$
(say $\si=0$) to correspond to the mouth, and attaching this
two-dimensional field theory to a
four-dimensional field theory (describing the AF region)
by sewing the line $\si=0$ onto the four-dimensional world line
of the black hole mouth. The details of this sewing are
in general complicated; the important point for our present purposes
is that after the $s$-wave fermion modes in the AF region enter
the throat, they become the massless two-dimensional $\psi$ modes.

In the AF region the dilaton field takes on the (nearly constant)
value $\phi_0$.
$\e^\phi$ governs
the strength of quantum effects in dilaton gravity.
If $\phi_0$ is sufficiently large and negative, then
quantum effects are
suppressed and there is little fermion backscattering until the fermion is
well into the throat where the dilaton grows linearly. Thus if
quantum backscattering does occur, for large negative ${\phi_0}$
it occurs mainly in the throat region where the dynamics are governed by
the two-dimensional effective action (\call{twoact}).

If the incoming fermion has sufficiently low energy, gravitational effects
(including backreaction) can be neglected, and (\call{twoact}) may
be approximated by
$$
S \approx \int d^2 \sigma\, \Bigl(i{\bar\psi}\Dsl\psi-{\txt {1\over 2}}
\e^{-2\phi(\si)}{F^2}\Bigr)\eqno(twoem)
$$
with $\phi(\sigma)$ given by (\call{ldv}).
The energies at which gravitational effects become important
will be discussed shortly.

The theory described by (\call{twoem}) is of course nothing but the
massless Schwinger model[\cite{SCHW}]
with a position dependent coupling constant:
$$
e(\sigma) = \e^{\phi(\si)} = \exp\bigl(\phi_0 + \sigma/2Q\bigr).\eqno(cplg)
$$
\noindent It may be bosonized by the usual procedure, which yields
$$
S = -\int  d^2 \sigma\, {1\over 2} \Bigl(
 (\nabla{b})^2 + {e^2\over2\pi}{b^2} \Bigr)
\eqno(bact)
$$
\noindent where following the conventions of [\cite{sidney}]
$\ep^{\mu\nu} {\partial_\mu}{b}
= \sqrt{\pi}\,{\bar\psi}{\gamma^\nu}\psi$.
This represents a free boson with mass $\mu=e(\sigma)/\sqrt{2\pi}$
which increases exponentially
with the distance into the throat. The equation of motion for ${b}$
$$
\square b = {1\over 2\pi}\exp\bigl(2\phi_0 + \sigma/Q \bigr) b.\eqno(beqn)
$$
\noindent can be solved exactly.\footnote{*}{For example by going into a
coordinate system for which it becomes ${\partial_+}{\partial_-}{b} = {b}$.}
However the scattering is easily understood qualitatively without doing
so. A fermion entering the throat at $\sigma = 0$
can be represented as some $b$ wave packet (see below).
For large negative ${\phi_0}$, the $b$ meson is
is effectively massless near the mouth, and so the wave packet
may propagate freely into the throat. However, as it gets further
down the throat the $b$ mass rises exponentially, and so the
wave packet must lose momentum.
The furthest it can possibly penetrate is to
$\si=\si_T$ such that
$$
{1\over\sqrt{2\pi}} \exp\bigl(\phi_0 + \sigma_T/2Q\bigr) = \omega\eqno(tpt)
$$
\noindent where $\omega$ is the incoming energy.
Before reaching $\si_T$ the wave packet
turns around and eventually exits from the throat region.
In general the shape of the outgoing wave packet is different from the
incoming one. When translated back into fermions, this means that a
single incoming fermion may backscatter into several outgoing ones.

Having established that a charged
massless fermion is always reflected by a magnetically charged
dilaton black hole, we now consider the case of a
fermion with mass $m$ in the AF region. In the throat, the
action now bosonizes to a massive Schwinger model
[\cite{sidney}]:
$$
S = -\int  d^2 \sigma\,  \biggl( \half
 (\nabla{b})^2 + {e^2\over 4\pi}{b^2}  - c m^2 \cos(2\sqrt{\pi}b)\biggr)
\eqno(msch)
$$
where $g$ depends on $\si$, just as in (\call{cplg}), and $c$ is
a numerical constant.
Let us assume that near the mouth the electromagnetic coupling
is very weak, $\exp(\phi_0)\ll m$, so that there is a region near
the mouth where $e^2(\si) \ll 8\pi^2 c m^2$. In this region
there are well-defined
kink solutions in the bosonized theory, which correspond to fermions
of the original theory.
If the fermion entering the throat is non-relativistic,
then by energy conservation
it will be unable to pair produce more fermions or bosons.
As it proceeds
down the throat it will have its
electric flux concentrated into a string behind it.
In the language of the bosonized theory this string
corresponds to the $b$ field being in the local potential minimum
at $b=\sqrt{\pi}$. At the location of the fermion there is a
kink, which takes the field to the global minimum $b=0$.
As the kink/fermion travels
down the throat not only does the string lengthen, but also
the electromagnetic coupling, and hence the string
tension, rises exponentially, and so the fermion must
eventually be pulled back out of the throat by its own flux string.

The situation is more complex for highly relativistic massive
fermions, because they may go beyond the region where
kink solutions exist, and then
the string can break, allowing a charged fermion to go back out of the throat
while neutral $b$-mesons continue in.
However
the $b$-meson mass rises as they go deeper (\call{beqn}),
and eventually, at $\si\le \si_T$ (\call{tpt}),
they too are reflected back out again.
Thus all the charge and energy of the incident fermion is reflected
back out of the throat, in a manner that is in principle computable
and described by a unitary $S$-matrix.

Several comments are
in order:
\item{1.} It was shown in [\cite{evan}] that gravitational effects
are important for this process when
$\exp(\phi_0+\sigma/2Q)$ is of
order one. Thus as long as $\omega\ll 1$, backscattering has
occurred before gravitational effects are important, and our neglect
in equation (\call{twoem})
is justified.
\item{2.} Some backscattering of fermions, dilatons or gravitons may
occur as the fermion enters the throat. This is hard to analyze in detail,
but it can be made small by decreasing ${\phi_0}$.
\item{3.} The case of ${N}$ flavors of electrons is both much more
interesting and much more difficult. The dynamics may be reduced to
the bosonized lagrangian
$$
{S = -\int d^2 \sigma\, \left[  \sum_{i=1}^N\,
\biggl(\half(\nabla
{b_i})^2-cm_i^2{\rm cos}(2{\sqrt \pi}b_i)\biggr) + {e^2(\sigma)\over 4\pi}
\biggl(\sum_{i=1}^N{b_i}\biggr)^2 \right] } \eqno(nact)
$$
\noindent where ${i}$ is a flavor index and we have assumed all flavors have
the same charge. For the special case of non-relativistic
massive fermions, the problem may be solved much as before: fermions
are pulled out of the throat by the string of electric flux produced in
their wake. However for the case of $N$ flavors of massless
fermions (which is similar to that of relativistic massive fermions),
the situation is more subtle.
In this case only the linear combination $\sum {b_i}$
of bosons acquires a position dependent mass from electromagnetic effects,
while ${N} - 1$ bosons remain
massless.
Thus not all components of a generic incoming wave packet will be
electromagnetically reflected. Those that are not will continue on until
$\exp\bigl(\phi_0 + \sigma/2Q\bigr)$ is of order one,
where gravitational effects
become important. The fascinating problem of gravitational backscattering in
this model has been
studied in [\cite{evan,ban,sus}].
\item{4.}In order to understand the interactions of fermions in the real
world
with charged black holes, one must consider
non-abelian gauge fields as in [\cite{rub, cal}]. Presumably our methods
can be extended to this case. One might also need to consider other types
(e.g. non-dilatonic) black holes. However, as emphasized in [\cite{rub,
cal}],
low-energy fermion $s$-wave scattering $is$ sensitive to short distance
physics.
Thus if there is a massless dilaton above some scale -- as suggested by
string theory -- it could affect the scattering of low-energy fermions from
small black holes.
\item{5.} We believe that the results found in this paper
have some bearing on the question
of ``catastrophic'' Hawking radiation in the $s$-wave mode.
This issue was raised in [\cite{hw}], where the
rate of mass loss for a black hole of
temperature $T$ through these modes was given as
$$
{dM\over dt}\Big|_{s=0} = \int { d\om \,\om\over
\e^{\omega/T} +1}~.\eqno(crrt)
$$
The transmission coefficient typically present in such expressions
was thought to be absent because, as argued above, the reflection coefficient
vanishes classically. The authors of [\cite{hw}] were concerned
that for ${a}>1$ extremal dilatonic black holes the temperature is
divergent, apparently leading to a divergent Hawking radiation rate
in $s$-wave modes. (Of course
very close to extremality this formula breaks down, since
the metric changes dramatically during the emission of a single
particle, so the arguments of [\cite{hw}] are only suggestive.)
However we have shown that (\call{crrt}) is an inaccurate formula
(even away
from extremality), since it
neglects the effect of electromagnetic back reaction. We have seen
that for the bosonized fermions these effects change
the reflection coefficient for the ${a}=1$ case from zero to one. If
a similar mechanism is operative in the ${a}>1$ case, as we believe it
should be, it will turn off radiation of $s$-wave modes near the extremal
limit.

It would be interesting to see if gravitational backreaction,
which is present for all particles, could have a similar effect of
backscattering all incident particles before they are swallowed by
an extremal black hole. If so, then particle-hole scattering would be
unitary.
Perhaps the calculations of the present paper could
serve as a simplified model for how that might actually occur.

\vskip .50truein
\centerline{ACKNOWLEDGEMENTS}

We are grateful to C.~Callan, S.~Giddings, J.~Harvey, and J.~Preskill
for useful conversations. This work was supported in part by DOE Grant
91ER40618 and NSF grant PHY-89-04035.

\references

\refis{ban}T. Banks, A. Dabholkar, M.R. Douglas, and M O'Loughlin, ``Are
horned particles the climax of Hawking evaporation?'' Rutgers preprint
RU-91-54.

\refis{sus}J.G. Russo, L. Susskind, and L. Thorlacius, ``Black hole
evaporation in 1+1 dimensions,'' Stanford preprint SU-ITP-92-4.

\refis{GIMA}G.W. Gibbons, ``Antigravitating black hole solitons with scalar
hair in N=4 supergravity,'' Nucl. Phys. {\bf B207} (1982) 337;
G.W. Gibbons and K. Maeda, ``Black holes and membranes in
higher-dimensional theories with dilaton fields,'' Nucl. Phys. {\bf B298}
(1988) 741.

\refis{alf} M. Alford, unpublished.

\refis{GHS}D. Garfinkle, G. Horowitz, and A. Strominger, ``Charged black holes
in string theory,'' Phys. Rev. {\bf D43} (1991) 3140.

\refis{HAW}S. W. Hawking, ``Particle creation by black holes,''
Comm. Math. Phys. {\bf 43} (1975) 199.

\refis{SCHW} J. Schwinger, Phys. Rev. 128 (1962) 2425;
J. Lowenstein and A. Swieca, Ann. Phys. 68 (1971) 172.

\refis{evan} C. Callan, S. Giddings, J. Harvey, and A. Strominger,
`` Evanescent black holes'', Phys. Rev. {\bf D45} (1992) 1005.

\refis{dynam} S. Giddings and A. Strominger, `` Dynamics of
extremal black holes'', UCSB preprint UCSB-TH-92-01.

\refis{hw} C. Holzhey and F. Wilczek, `` Black holes as elementary
particles'', IAS preprint IASSNS-HEP-91/71.

\refis{sidney} S. Coleman, Ann. Phys. 101 (1976) 239;
S. Coleman, R. Jackiw, L. Susskind, Ann. Phys. 93 (1975) 267.

\refis{cal} C.~Callan, Phys. Rev. D25 (1982) 2141; Phys. Rev.
D26 (1982) 2058; Nucl. Phys. B212 (1983) 391.

\refis{rub} V. Rubakov, Pis'ma Zh. Eksp. Teor. Fiz. 33 (1981) 658 (JETP
 Lett. 33 (1981) 644); Nucl. Phys. B203 (1982) 311.

\endreferences

\endit